\documentclass[preprint,amsmath,amssymb,groupedaddress]{revtex4}
\usepackage{color}
\usepackage{graphicx}
\usepackage{dcolumn}
\usepackage{bm}
\usepackage{amsmath}
\usepackage{epsfig}
\begin{document}
\draft
\title{Fidelity and entanglement close to quantum phase transition in a two-leg $XXZ$ spin ladder}
\date{\today}
\author{Jie Ren$^{1,2}$\footnotetext{E-mail: jren01@163.com}}
\author{Shiqun Zhu$^2$\footnote{Corresponding author, E-mail: szhu@suda.edu.cn}}

\affiliation{$^1$Department of Physics and Jiangsu Laboratory of Advanced
Functional Material, Changshu Institute of Technology, Changshu, Jiangsu 215500, People's Republic of China \\
$^2$School of Physical Science and Technology, Suzhou University,
Suzhou, Jiangsu 215006, People's Republic of China}

\begin{abstract}

The fidelity susceptibility and entanglement entropy in a system of
two-leg $XXZ$ spin ladder with rung coupling is investigated by
using exact diagonalization of the system. The effects of rung
coupling on fidelity susceptibility, entanglement entropy and
quantum phase transition are analyzed. It is found that the quantum
phase transition between two different $XY$ phases can be well
characterized by the fidelity susceptibility. Though the quantum
phase transition from $XY$ phase to rung singlet phase can be hardly
detected by fidelity susceptibility, it can be predicted by the
first derivative of the entanglement entropy of the system.

\vskip 0.6 cm

\end{abstract}

\maketitle

\section{Introduction}

The quantum phase transition has attracted much attention in
low-dimensional quantum systems. It implies fluctuations which
happened at zero temperature ~\cite{Sachdev}. When a controlling
parameter changes across critical point, some properties of the
many-body system will change dramatically. Many results show that
entanglement exists naturally in the spin chain when the temperature
is at zero. The quantum entanglement of a many-body system has been
paid much attention since the entanglement is considered as the
heart in quantum information and computation
~\cite{Nielsen,Bennett}. As the bipartite entanglement measurement
in a pure state, the von Neumann entropy ~\cite{Bennett1} in the
antiferromagnetic anisotropic and isotropic spin chains
~\cite{Vidal,Latorre} were investigated respectively. By using the
quantum many-body theory and quantum-information theory, von Neumann
entropy was applied to detect quantum critical behaviors
~\cite{Preskill,Amico,Gu01,Kitaev}. Meanwhile, the ground state
fidelity was used to qualify quantum phase transition in the last
few years
~\cite{Zhou01,Quan,Gu1,Abasto,Oelkers,Cozzini,Venuti,Buonsante,You,Yang,Zhou02,Tzeng,Ren,Ren01}.
It is shown that the fidelity and the entanglement entropy have
similar predictive power for identifying quantum phase transitions
in the spin systems.

Spin ladders are examples of interacting many-body systems which
exhibit many novel phenomena. The spin ladders exist in the
$Cu_2O_3$ subsystem of the compound $Sr_14Cu_24O_3$ ~\cite{Ueda} and
real crystals $(C_6H_11NH_3)CuBr_3(CHAB)$ ~\cite{K}. Recently, Bose
\emph{it el} ~\cite{Bose} studied the fidelity, the entanglement and
the quantum phase transition of a spin-$1/2$ antiferromagnetic
Heisenberg spin ladder in an external magnetic field. The effects of
magnetic field on variation of fidelity and entanglement measures
were investigated. It was found that the variation of fidelity and
entanglement measured points close to the quantum criticality. It
would be interesting to investigate the effects of rung interaction
on the variation of the fidelity, the entanglement and the quantum
phase transition.

In this paper, the fidelity and entanglement in a two-leg $XXZ$ spin
ladder system is investigated using exact diagonalization. In
section II, the Hamiltonian of the two-leg $XXZ$ spin ladder system
is presented. In section III, the effect of rung interaction on
ground state fidelity is investigated. Its relation with quantum
phase transition is analyzed. The effect of rung interaction on the
entanglement entropy is also calculated and analyzed in section IV.
A discussion concludes the paper.

\section{Hamiltonian of Spin Ladder}

The Hamiltonian of a two-leg antiferromagnetic Heisenberg spin
ladder is given by

\begin{equation}
\label{eq1}
H=H^1_{leg}+H^2_{leg}+H_{rung},\\
\end{equation}
where the Hamiltonian $H^{\alpha}_{leg}$ for leg $\alpha$
($\alpha=1$ or $2$) is given by

\begin{equation}
\label{eq2}
H^{\alpha}_{leg}=\sum_{i=1}^{N}J(S^x_{\alpha,i}S^x_{\alpha,i+1}+S^y_{\alpha,i}S^y_{\alpha,i+1}+\Delta S^z_{\alpha,i}S^z_{\alpha,i+1}),\\
\end{equation}
and the inter-leg coupling is given by

\begin{equation}
\label{eq3}H_{rung}=\sum_{i=1}^{N}J_{rung}(S^x_{1,i}S^x_{2,i}+S^y_{1,i}S^y_{2,i}+S^z_{1,i}S^z_{2,i}),\\
\end{equation}
where $S^{x, y, z}_{\alpha,j}$ are spin operators on the $i$-th
rung, the index $\alpha=1,2$ denotes the leg in the ladder, $N$ is
the length of the spin ladder, $J>0$ denotes the antiferromagnetic
coupling, $\Delta$ is anisotropic interaction, $J_{rung}$ denotes
rung interaction. The schematic diagram of the two-leg spin ladder
is shown in Fig. 1. In the paper, the opened boundary condition is
considered, $J=1$ and $\Delta=-0.5$ are chosen for simplicity. It is
predicted that a novel $XY2$ phase $J_{rung}^{C1}< J_{rung}<
J_{rung}^{C2}$ appears between the $XY1$ phase and the rung singlet
phase. The XY phase belongs to the universality class of
Tomonaga-Luttinger liquid. A number of numerical studies have shown
that phase transition from $XY1$ to $XY2$ occurs at $J_{rung}^{C1} =
0$, and phase transition from $XY2$ to  rung singlet phase point
$J_{rung}^{C2}$ depends on the anisotropy interaction in the legs.
When $\Delta=-0.5$, the$J_{rung}^{C2}=0.373$\cite{Vekua,Hijii}.

\section{Fidelity Susceptibility}

The ground state fidelity and the fidelity susceptibility can be
applied to detect the existence of the quantum phase transition. A
general Hamiltonian of a quantum many-body system can be written as
$ H(\lambda)=H_0+\lambda H_I $ where $H_I$ is the driving
Hamiltonian and $\lambda$ denotes its strength. If $\rho(\lambda)$
represents a state of the system $H(\lambda)$, the fidelity between
states $\rho(\lambda)$ and $\rho(\lambda+\delta)$ can be defined as

\begin{equation}
\label{eq4}F(\lambda,\delta)=Tr[\sqrt{\rho^{1/2}(\lambda)\rho(\lambda+\delta)\rho^{1/2}(\lambda)}].
\end{equation}
If the state can be written as $\rho=|\psi\rangle\langle \psi|$, Eq.
(4) can be written as $F(\lambda,\delta)=|\langle
\psi(\lambda)|\psi(\lambda+\delta)\rangle|$. Because
$F(\lambda,\delta)$ reaches its maximum value $F_{max}=1$ at
$\delta=0$, on expanding the fidelity in powers of $\delta$, the
first derivative $\frac{\partial F(\lambda,\delta=0)}{\partial
\lambda}=0$. Then the fidelity can written by

\begin{equation}
\label{eq5}F(\lambda,\delta)\simeq1+\frac{\partial^2F(\lambda,\delta)}{2\partial\lambda^2}|_{\lambda=\lambda'}\delta^2,
\end{equation}
where $lambda'=lambda$ should probably read as delta=0. The average
fidelity susceptibility $S(\lambda,\delta)$ can be given by
~\cite{Cozzini,Buonsante}

\begin{equation}
\label{eq6}S(\lambda,\delta)=\lim_{\delta\rightarrow
0}\frac{2[1-F(\lambda,\delta)]}{N\delta^2}.
\end{equation}

For models that are not exactly solvable, the exact diagonalization
can be used to obtain the ground state. The fidelity $F$ and the
fidelity susceptibility $S$ is calculated and plotted in Fig. 2 as a
function of anisotropy parameter $J_{rung}$ for different sizes. The
parameters are chosen as $N=8, 10, 12$ and $\delta=0.001$
~\cite{Tzeng}. In Fig. 2(a), there is a valley in the fidelity. The
locations of the minimal value for all sizes at $J_{rung}^{C1}=0$.
In Fig. 2(b), one peak locates at the same point as the valley in
Fig. 2(a). It means that the fidelity and fidelity susceptibility
can predict the quantum phase transition between $XY1$ and $XY2$
phases. In the region of $J_{rung}<0$, $\langle S_{1,j}^x
S_{2,j}^x\rangle >0$ is obtained from the Marshall-Lieb-Mattis
theorem. Meanwhile, in the region of $J_{rung}>0$, $\langle
S_{1,j}^x S_{2,j}^x\rangle <0$. These two $XY$ phases have different
symmetry~\cite{Nomura,Tomonaga}. So the fidelity and fidelity
susceptibility can predict the quantum phase transition point.
However, it is difficult to detect the phase transition between
$XY2$ and rung singlet phases since the phase transition between
$XY2$ and rung singlet phases is Berezinskii-Kosterlitz-Thouless
(BKT) type. The fidelity is feeble to characterize the
Berezinskii-Kosterlitz-Thouless (BKT) type phase transition
~\cite{Chen,Chen01}.

\section{Entanglement Entropy}

Similarly, the ground state entanglement can also be used to detect
the quantum phase transition. The entropy can be chosen as a
measurement of the pairwise entanglement. The entropy can be defined
as follows. Let $\rho_{AB}$ be the ground state of a chain of $N$
qubits, the reduced density matrix of part A can be written as
$\rho_{A}=Tr_{B}\rho_{AB}$. The bipartite entanglement between parts
$A$ and $B$ can be measured by the entanglement entropy as

\begin{equation}
\label{eq7}E_{AB}=-Tr(\rho_{A(B)}\log_2\rho_{A(B)}).
\end{equation}

By using the method of exact diagonalization, the entropy of the
ground state can be calculated. The entropy $E_{rung}$ and the first
derivative $dE_{rung}/dJ_{rung}$ of the entropy is plotted as a
function of the rung interaction $J_{rung}$ in Fig. 3 with different
size of $N=8, 10, 12$. In Fig. 3(a), the entanglement entropy
between central rung and rest of the system is plotted. There is a
peak in $E_{rung}$ at $J_{rung}^{C1}=0$. This means that the quantum
phase transition between two different $XY$ phases can be well
characterized by entropy, while the transition from $XY2$ to rung
singlet phases can be hardly detected by the entropy. It seems that
this is due to the monogamy property of the
entropy\cite{Amico,Ren01}. In Fig. 3(b), the first derivative of the
entropy is plotted. There is a sharp change in $dE_{rung}/dJ_{rung}$
at $J_{rung}^{C1}=0$. This means that the quantum phase transition
between two different $XY$ phases occurs. There is also a peak in
$dE_{rung}/dJ_{rung}$ near $J_{rung}^{C2}=0.37$. This peak indicates
the quantum phase transition from $XY2$ phase to rung singlet phase.
This is in consistent with the result of the calculation of the
correlation function. It was shown that the phase transition from
$XY1$ to $XY2$ occurred at $J_{rung}^{C1} = 0$. The phase transition
from $XY2$ to rung singlet phase appeared at $J_{rung}^{C2}=0.373$
when $\Delta=-0.5$ ~\cite{Vekua,Hijii}. It is clear that the
critical properties can be captured by the first derivative of the
entropy as a function of the rung interaction $J_{rung}$
~\cite{Amico,Osterloh,Liu,Ren01,Wu}.

The entropy of the diagonal two qubits of the central rung is
plotted in Fig. 4. There is a peak at $J_{rung}^{C1}=0$ and a valley
near $J_{rung}^{C2}=0.37$. It is clear that the quantum phase
transition between $XY1$ and $XY2$ phases appears at
$J_{rung}^{C1}=0$ while the transition from $XY2$ phase to rung
singlet phase occurs near $J_{rung}^{C2}=0.37$. It seems that the
entropy of the diagonal two qubits of the central rung can also
predict the two different kinds of quantum phase transitions in a
two-leg ladder system.

\section{Discussion}

The fidelity susceptibility and the entanglement entropy of a
two-leg $XXZ$ spin ladder system are studied numerically. By using
the exact diagonalization, the effect of rung coupling on fidelity
susceptibility and entanglement entropy is investigated. Their
relations with quantum phase transition are analyzed. It is found
that the quantum phase transition between two different $XY$ phases
can be well characterized by fidelity susceptibility, while the
transition from $XY2$ to rung singlet phases can be hardly detected
by the fidelity susceptibility. The first derivative of the pairwise
entanglement entropy and the entropy between diagonal two qubits of
the central rung and the rest of the system can detect the quantum
phase transition between two different $XY$ phases and the
transition from $XY2$ to rung singlet phases.

\vskip 0.6 cm

{\textbf{Acknowledgments}}

It is a pleasure to thank Yinsheng Ling, Jianxing Fang and Xiang Hao
for their many helpful discussions. The financial support from the
National Natural Science Foundation of China (Grant No. 10774108) is
gratefully acknowledged.

\newpage

\begin{center}{\Large \bf Figure Captions}\end{center}

\textbf{Fig. 1}

The schematic diagram of a two-leg spin ladder.

\textbf{Fig. 2}

The fidelity $F$ and the fidelity susceptibility $S$ is plotted as a
function of the interaction $J_{rung}$ for different size $N$. (a).
The fidelity $F$. (b). The fidelity susceptibility $S$.

\textbf{Fig. 3}

The entropy $E_{rung}$ and the first derivative
$dE_{rung}/dJ_{rung}$ of the entropy between two qubits of central
rung is plotted as a function of the interaction $J_{rung}$ for
different size $N$. (a). The entropy $E_{rung}$. (b). The first
derivative $dE_{rung}/dJ_{rung}$ of the entropy.

\textbf{Fig. 4}

The entropy of the diagonal two qubits of central rung is plotted as
a function of the interaction $J_{rung}$ for different size $N$.

\newpage
\centerline{\epsfig{file=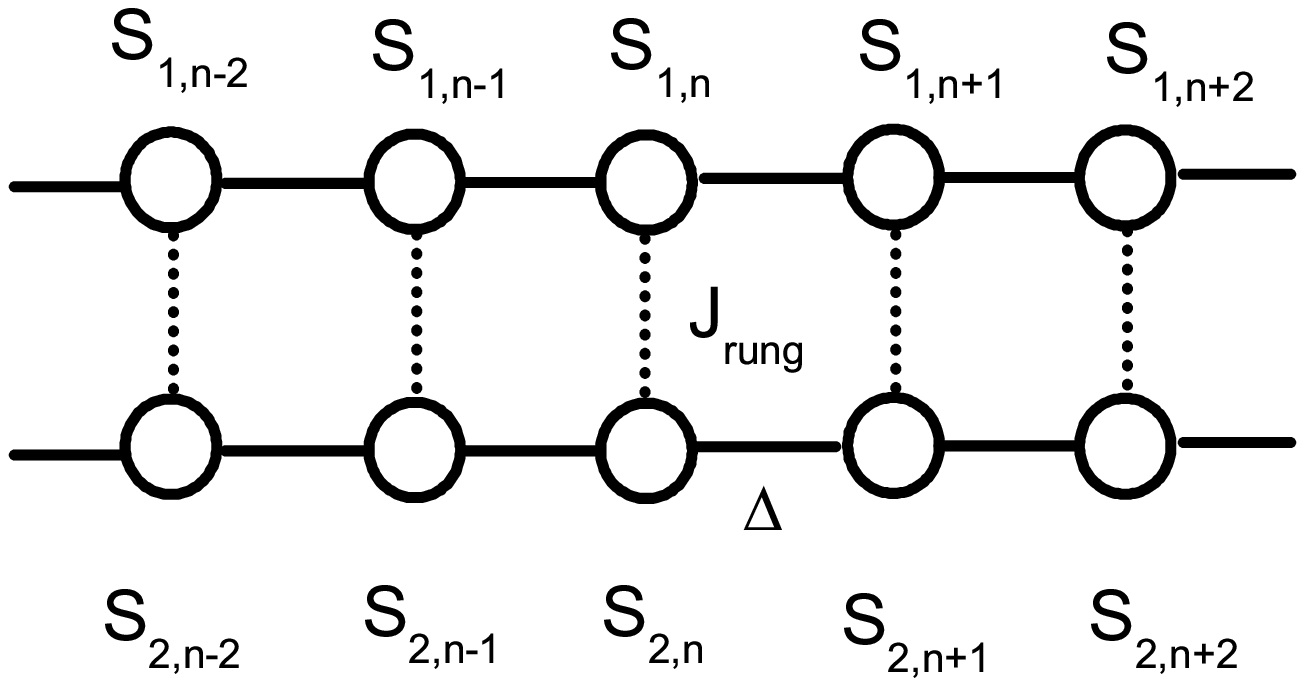,width=300pt}} \centerline{\bf
\large Fig. 1}

\newpage

\centerline{\epsfig{file=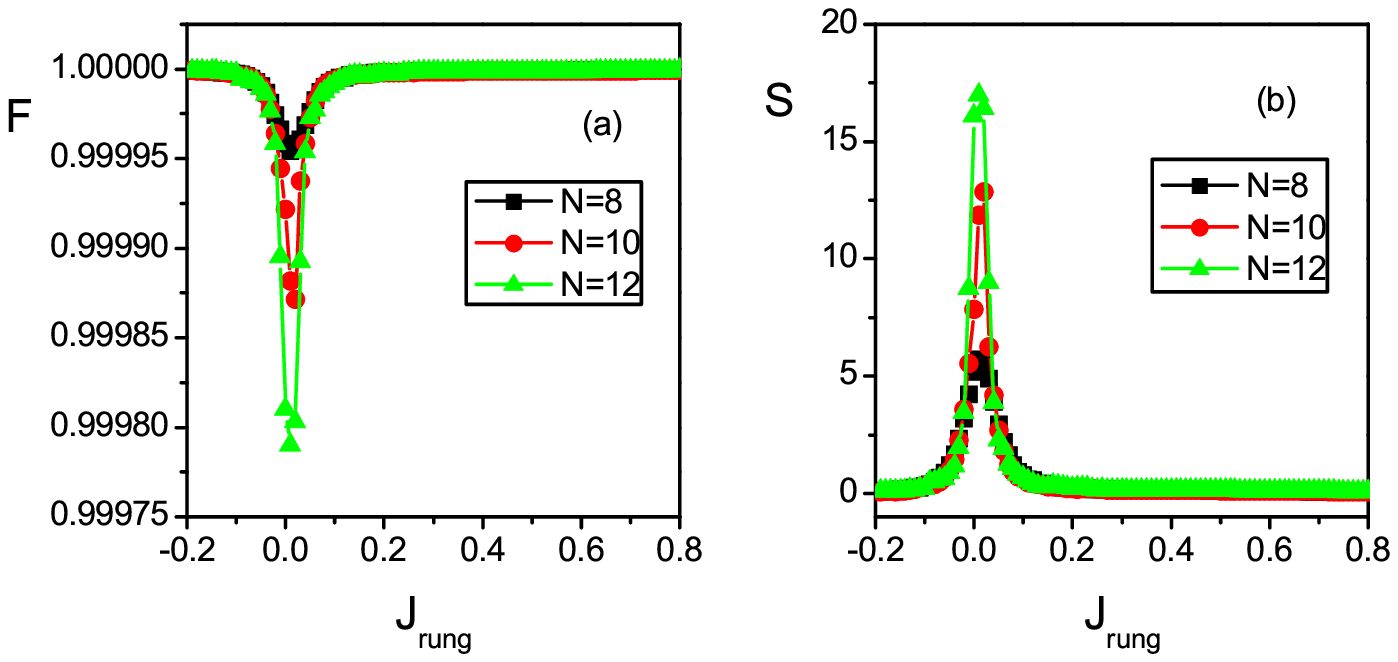,width=500pt}} \centerline{\bf
\large Fig. 2}

\newpage
\centerline{\epsfig{file=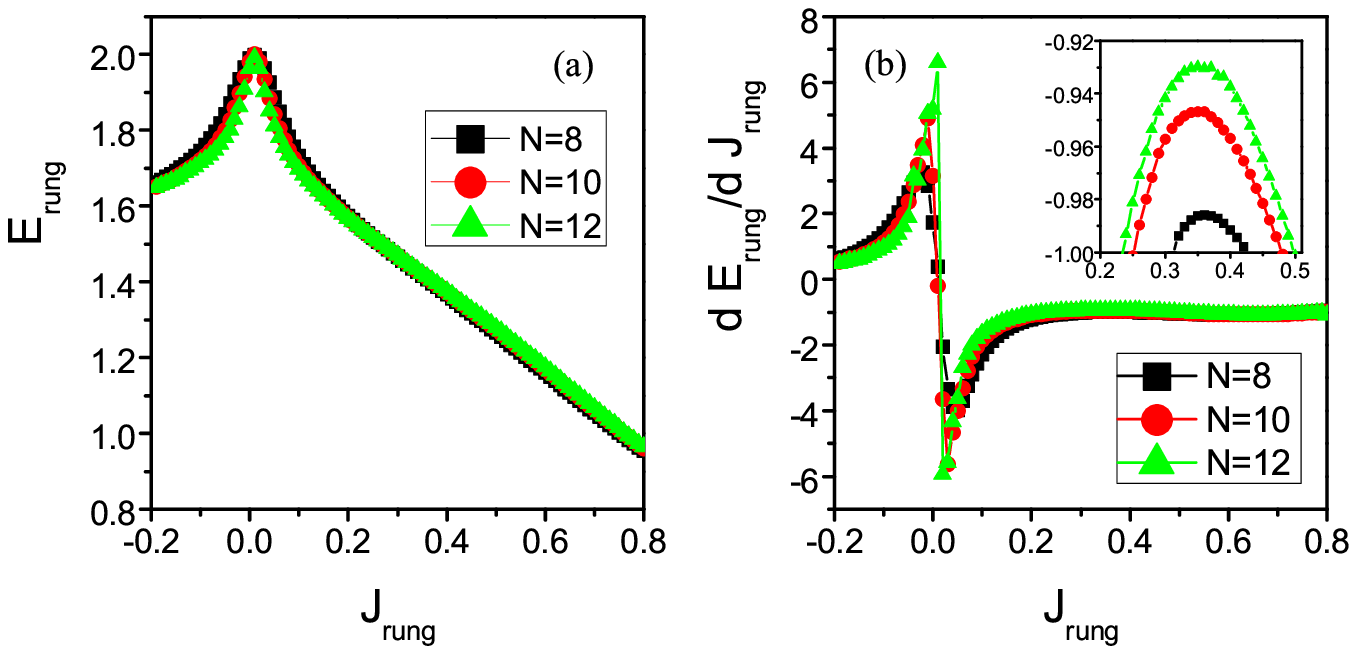,width=500pt}} \centerline{\bf
\large Fig. 3}

\newpage

\centerline{\epsfig{file=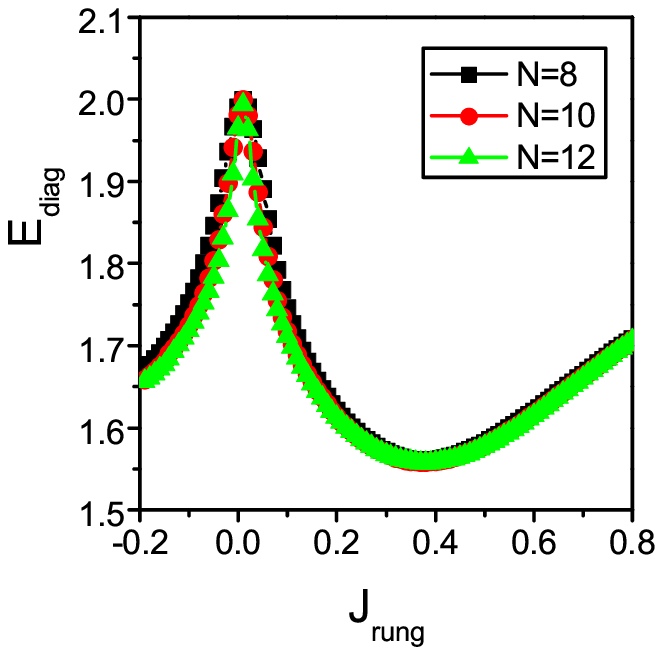,width=300pt}} \centerline{\bf
\large Fig. 4}

\end{document}